\documentclass{ws-ijmpcs}

\begin{document}

\markboth{H.~T.~CHO, A.~S.~CORNELL, JASON~DOUKAS, AND WADE~NAYLOR}
{ANGULAR EIGENVALUES...}

%
\catchline{}{}{}{}{}
%

\title{ANGULAR EIGENVALUES OF HIGHER-DIMENSIONAL KERR-(A)dS BLACK HOLES WITH TWO ROTATIONS
}

\author{H.~T.~CHO}

\address{Department of Physics, Tamkang University\\
Tamsui, New Taipei City, Taiwan\\
htcho@mail.tku.edu.tw}

\author{A.~S.~CORNELL}

\address{National Institute for Theoretical Physics; School of Physics,
University of the Witwatersrand\\
Wits 2050, South Africa\\
alan.cornell@wits.ac.za}

\author{JASON~DOUKAS}

\address{Yukawa Institute for Theoretical Physics, Kyoto University\\
Kyoto, 606-8502, Japan\\
jasonad@yukawa.kyoto-u.ac.jp}

\author{WADE~NAYLOR}

\address{International College \& Department of Physics, Osaka University\\
Toyonaka, Osaka 560-0043, Japan\\
naylor@het.phys.sci.osaka-u.ac.jp}

\maketitle

\begin{history}
\received{Day Month Year}
\revised{Day Month Year}
\end{history}

\begin{abstract}
In this paper, following the work of Chen, L\"u and Pope, we present the general metric for Kerr-(A)dS black holes with two rotations. The corresponding Klein-Gordon equation is separated explicitly, from which we develop perturbative expansions for the angular eigenvalues in powers of the rotation parameters with $D\geq 6$.

\end{abstract}

\ccode{PACS numbers: 02.30.Mv, 04.50.-h, 04.70.-s, 11.25.-w}

\section{Introduction}\label{sec:intro}

After the advent of the brane world scenario \cite{randall} and the AdS/CFT correspondence~\cite{maldacena}, there has been an increase in interest in the study of higher-dimensional black holes. The Kerr metric was first generalized to higher dimensions in the seminal paper by Myers and Perry \cite{myers}. One of the unexpected results is that for some rotating black holes event horizons exist for arbitrarily large values of the rotation parameters. The stability of such black holes are certainly in question~\cite{emparan}. These are asymptotically flat black holes. The first asymptotically non-flat higher-dimensional Kerr metric was given by Hawking {\it et al}.~\cite{hawking} for five dimensions. Subsequent generalizations to arbitrary dimensions were done by Gibbons {\it et al}.~\cite{gibbons}, finally resulting in the most general Kerr-(A)dS-NUT metric by Chen, L\"u and Pope~\cite{chen}.

The study of the wave equations in these black hole spacetimes was initiated by Frolov and Stojkovic~\cite{frolovst} by analyzing the Klein-Gordon equation in five dimensions. The analysis relied crucially on the method of the separation of variables. The problem of separability of these wave equations in higher dimensions is a difficult one. Even for the Klein-Gordon case, early attempts were only aimed at special cases~\cite{vasudevan}. Finally, using the Chen-L\"u-Pope metric, Frolov, Krtous and Kubiznak~\cite{frolovkk} were able to separate the geodesic equation and the Klein-Gordon equation in the most general setting.

However, the research on the scalar wave equation so far is focused mostly on the case with one rotation, the so-called simply-rotating case. Notably, the stability of the scalar perturbation in six and higher dimensions was considered in the ultra-rotating cases~\cite{morisawa,cardoso,kodamakz1}. Apparently no instability was found. Then, due to the interests in AdS spacetimes, these considerations were extended to Kerr-AdS black holes~\cite{kodamakz2}. Here the expected superradiant instability did indeed show up. In addition the Hawking radiation in these spacetimes~\cite{doukas,kantikkpz} are calculated with possible application to the production and decay of LHC black holes~\cite{kanti}.

In this paper we would like to explore the cases with two rotations. In the next section we present the general metric of the Kerr-(A)dS black hole with two rotations. The corresponding Klein-Gordon equation will be separated into one radial equation and two angular ones. In Section 3, we shall develop perturbative expansions of the angular eigenvalues in powers of the rotation parameters. Conclusions and discussions are given in Section 4.

\section{The metric and the separated equations of the Klein-Gordon equation with two rotations}\label{sec:KG}

In this section we shall first present the Kerr-(A)dS metric with two rotations. We start with the metric for $D=2n$, that is, for even dimensions. However, the result we obtain in the end should also be valid for odd dimensional cases. For $D=2n$, there are at most $n-1$ rotation directions so we have $a_{i}$, $i=1,2,\dots,n-1$. This metric, satisfying $R_{\mu\nu}=-3g^{2}g_{\mu\nu}$, can be expressed as follows \cite{chen}.
\begin{eqnarray}
ds^{2}&=&\frac{U}{X}dr^{2}+\sum_{\alpha=1}^{n-1}\frac{U_{\alpha}}{X_{\alpha}}dy_{\alpha}^{2}
-\frac{X}{U}\left[Wd\tilde{t}-\sum_{i=1}^{n-1}\gamma_{i}d\tilde{\phi}_{i}\right]^{2}\nonumber\\
&&\ \ \sum_{\alpha=1}^{n-1}\frac{X_{\alpha}}{U_{\alpha}}\left[\frac{(1+g^{2}r^{2})W}{1-g^{2}y_{\alpha}^{2}}d\tilde{t}
-\sum_{i=1}^{n-1}\frac{(r^{2}+a_{i}^{2})\gamma_{i}}{a_{i}^{2}-y_{\alpha}^{2}}d\tilde{\phi}_{i}\right]^{2},\label{metric}
\end{eqnarray}
where
\begin{eqnarray}
U&=&\prod_{\alpha=1}^{n-1}(r^{2}+y_{\alpha}^{2}),\ \ \ \ \ \ U_{\alpha}=-(r^{2}+y_{\alpha}^{2})\prod_{\beta=1,\beta\neq\alpha}^{n-1}(y_{\beta}^{2}-y_{\alpha}^{2}),\\
X&=&(1+g^{2}r^{2})\prod_{k=1}^{n-1}(r^{2}+a_{k}^{2})-2Mr,\ \ \ \ \ \ X_{\alpha}=-(1-g^{2}y_{\alpha}^{2})\prod_{k=1}^{n-1}(a_{k}^{2}-y_{\alpha}^{2}),\\
W&=&\prod_{\alpha=1}^{n-1}(1-g^{2}y_{\alpha}^{2}),\ \ \ \ \ \ \gamma_{i}=\prod_{\alpha=1}^{n-1}(a_{i}^{2}-y_{\alpha}^{2}),\\
t&=&\tilde{t}\prod_{i=1}^{n-1}(1-g^{2}a_{i}^{2}),\ \ \ \ \ \ \phi_{i}=a_{i}(1-g^{2}a_{i}^{2})\tilde{\phi}_{i}\prod_{k=1,k\neq i}^{n-1}(a_{i}^{2}-a_{k}^{2}),
\end{eqnarray}
with $1\leq\alpha,i\leq n-1$. $\phi_{i}$ is the azimuthal angle for each $a_{i}$. With the direction cosines $\mu_{i}$, $i=1,\dots,n$, the metric for a unit $D-2$ sphere is just
\begin{equation}
d\Omega^{2}=\sum_{i=1}^{n}d\mu_{i}^{2}+\sum_{i=1}^{n-1}\mu_{i}^{2}d\phi_{i}^{2},
\end{equation}
subject to the constraint $\sum_{i=1}^{n}\mu_{i}^{2}=1$. This constraint can be solved in terms of the unconstrained latitude variables $y_{\alpha}$'s,
\begin{equation}
\mu_{i}^{2}=\frac{\prod_{\alpha=1}^{n-1}(a_{i}^{2}-y_{\alpha}^{2})}{a_{i}^{2}\prod_{k=1,k\neq i}^{n-1}(a_{i}^{2}-a_{k}^{2})},\ \ \ \ \ \mu_{n}^{2}=\frac{\prod_{\alpha=1}^{n-1}y_{\alpha}^{2}}{\prod_{i=1}^{n-1}a_{i}^{2}}.\label{defy}
\end{equation}
The metric for the unit sphere can then be written as
\begin{equation}
d\Omega^{2}=\sum_{\alpha=1}^{n-1}g_{\alpha}dy_{\alpha}^{2}+\sum_{i=1}^{n-1}\mu_{i}^{2}d\phi_{i}^{2},\label{spheremetric}
\end{equation}
with
\begin{equation}
g_{\alpha}=\frac{\prod_{\beta=1,\beta\neq\alpha}^{n-1}(y_{\beta}^{2}-y_{\alpha}^{2})}{\prod_{k=1}^{n-1}(a_{k}^{2}-y_{\alpha}^{2})}.
\end{equation}

Now, to obtain a general metric with two rotations, we take the limit $a_{3}$, $a_{4}$, $\dots$, $a_{n}\rightarrow 0$ while assuming that $a_{1}>a_{2}>a_{3}>\cdots>a_{n-1}$. From the definition in Eq.~(\ref{defy}) we see that $y_{i}$ is of the same order of magnitude as $a_{i}$. With this in mind, we take the limit to arrive at the metric with two rotations for general dimensions $D\geq 6$,
\begin{eqnarray}
ds^{2}&=&-\frac{\Delta_{r}}{(r^{2}+y_{1}^{2})(r^{2}+y_{2}^{2})}
\left[\frac{(1-g^{2}y_{1}^{2})(1-g^{2}y_{2}^{2})}{(1-g^{2}a_{1}^{2})(1-g^{2}a_{2}^{2})}dt-\frac{(a_{1}^{2}-y_{1}^{2})(a_{1}^{2}-y_{2}^{2})}
{(1-g^{2}a_{1}^{2})(a_{1}^{2}-a_{2}^{2})}\frac{d\phi_{1}}{a_{1}}\right.\nonumber\\
&&\ \ \ \ \ \ \ \ \ \ \ \ \ \ \ \ \ \ \ \ \ \ \ \ \ \ \ \ \ \ \ \ \ \ \ \ \ \left.-\frac{(a_{2}^{2}-y_{1}^{2})(a_{2}^{2}-y_{2}^{2})}
{(1-g^{2}a_{2}^{2})(a_{2}^{2}-a_{1}^{2})}\frac{d\phi_{2}}{a_{2}}\right]^{2}\nonumber\\
&&+\frac{\Delta_{y_{1}}}{(r^{2}+y_{1}^{2})(y_{2}^{2}-y_{1}^{2})}
\left[\frac{(1+g^{2}r^{2})(1-g^{2}y_{2}^{2})}{(1-g^{2}a_{1}^{2})(1-g^{2}a_{2}^{2})}dt-\frac{(r^{2}+a_{1}^{2})(a_{1}^{2}-y_{2}^{2})}
{(1-g^{2}a_{1}^{2})(a_{1}^{2}-a_{2}^{2})}\frac{d\phi_{1}}{a_{1}}\right.\nonumber\\
&&\ \ \ \ \ \ \ \ \ \ \ \ \ \ \ \ \ \ \ \ \ \ \ \ \ \ \ \ \ \ \ \ \ \ \ \ \ \left.-\frac{(r^{2}+a_{2}^{2})(a_{2}^{2}-y_{2}^{2})}
{(1-g^{2}a_{2}^{2})(a_{2}^{2}-a_{1}^{2})}\frac{d\phi_{2}}{a_{2}}\right]^{2}\nonumber\\
&&+\frac{\Delta_{y_{2}}}{(r^{2}+y_{2}^{2})(y_{1}^{2}-y_{2}^{2})}
\left[\frac{(1+g^{2}r^{2})(1-g^{2}y_{1}^{2})}{(1-g^{2}a_{1}^{2})(1-g^{2}a_{2}^{2})}dt-\frac{(r^{2}+a_{1}^{2})(a_{1}^{2}-y_{1}^{2})}
{(1-g^{2}a_{1}^{2})(a_{1}^{2}-a_{2}^{2})}\frac{d\phi_{1}}{a_{1}}\right.\nonumber\\
&&\ \ \ \ \ \ \ \ \ \ \ \ \ \ \ \ \ \ \ \ \ \ \ \ \ \ \ \ \ \ \ \ \ \ \ \ \ \left.-\frac{(r^{2}+a_{2}^{2})(a_{2}^{2}-y_{1}^{2})}
{(1-g^{2}a_{2}^{2})(a_{2}^{2}-a_{1}^{2})}\frac{d\phi_{2}}{a_{2}}\right]^{2}\nonumber\\
&&+\frac{(r^{2}+y_{1}^{2})(r^{2}+y_{2}^{2})}{\Delta_{r}}dr^{2}+\frac{(r^{2}+y_{1}^{2})(y_{2}^{2}-y_{1}^{2})}{\Delta_{y_{1}}}dy_{1}^{2}
+\frac{(r^{2}+y_{2}^{2})(y_{1}^{2}-y_{2}^{2})}{\Delta_{y_{2}}}dy_{2}^{2}\nonumber\\
&&+r^{2}\left(\frac{y_{1}^{2}y_{2}^{2}}{a_{1}^{2}a_{2}^{2}}\right)d\Omega^{2}_{D-6},
\label{kerradsmetric}
\end{eqnarray}
and
\begin{eqnarray}
\Delta_{r}&=&(1+g^{2}r^{2})(r^{2}+a_{1}^{2})(r^{2}+a_{2}^{2})-2Mr^{7-D},\\
\Delta_{y_{1}}&=&(1-g^{2}y_{1}^{2})(a_{1}^{2}-y_{1}^{2})(a_{2}^{2}-y_{1}^{2}),\\
\Delta_{y_{2}}&=&(1-g^{2}y_{2}^{2})(a_{1}^{2}-y_{2}^{2})(a_{2}^{2}-y_{2}^{2}).
\end{eqnarray}
Here we have kept the variables $y_{1}$ and $y_{2}$ instead of writing them in terms of angle variables. This is because the relationship between them, as shown in Eq.~(\ref{defy}), is a bit complicated to write out explicitly. If we solve $y_{1}$ and $y_{2}$ in terms of $\mu_{1}$ and $\mu_{2}$, we have
\begin{eqnarray}
y_{1,2}^{2}&=&\frac{1}{2}\Bigg[a_{1}^{2}(1-\mu_{1}^{2})+a_{2}^{2}(1-\mu_{2}^{2})\nonumber\\
&&\ \ \ \ \ \ \ \ \ \ \ \ \pm\sqrt{4a_{1}^{2}a_{2}^{2}(\mu_{1}^{2}+\mu_{2}^{2}-1)
+(a_{1}^{2}(1-\mu_{1}^{2})+a_{2}^{2}(1-\mu_{2}^{2}))^{2}}\Bigg].
\end{eqnarray}
However, it is important to note that $y_{1}$ and $y_{2}$ must be constrained to be
\begin{eqnarray}
a_{2}\leq y_{1}\leq a_{1}\ \ \ \ \ ;\ \ \ \ \ 0\leq y_{2}\leq a_{2}
\end{eqnarray}
in order to have Eq.~(\ref{defy}) well-defined.

The separation of the Klein-Gordon equation in the general Kerr-(A)dS metric has been achieved by Frolov {\it et al}.~\cite{frolovkk}. Here we shall present explicitly the separated equation for the case with two rotations. Writing the Klein-Gordon field as
\begin{eqnarray}
\Phi=R_{r}(r)R_{y_{1}}(y_{1})R_{y_{2}}(y_{2})e^{i\psi_{0}\Psi_{0}}e^{i\psi_{1}\Psi_{1}}e^{i\psi_{2}\Psi_{2}}Y(\Omega),
\end{eqnarray}
where
\begin{eqnarray}
\Psi_{0}&=&-\omega+g^{2}(m_{1}a_{1}+m_{2}a_{2}),\\
\Psi_{1}&=&-\omega(a_{1}^{2}+a_{2}^{2})+m_{1}a_{1}(1+g^{2}a_{2}^{2})+m_{2}a_{2}(1+g^{2}a_{1}^{2}),\\
\Psi_{2}&=&-\omega a_{1}^{2}a_{2}^{2}+m_{1}a_{1}a_{2}^{2}+m_{2}a_{1}^{2}a_{2},
\end{eqnarray}
the Klein-Gordon equation $\partial_{\mu}\left(\sqrt{-g}g^{\mu\nu}\partial_{\nu}\Phi\right)=0$ can be simplified to the radial equation
\begin{eqnarray}
&&\frac{1}{r^{D-6}}\frac{d}{dr}\left(r^{D-6}\Delta_{r}\frac{dR_{r}}{dr}\right)+\Bigg[\frac{(r^{2}+a_{1}^{2})^{2}(r^{2}+a_{2}^{2})^{2}\omega^{2}}
{\Delta_{r}}\nonumber\\
&&\ \ \ \ \ -\frac{2\omega(1+g^{2}r^{2})(r^{2}+a_{1}^{2})^{2}(r^{2}+a_{2}^{2})^{2}}{\Delta_{r}}
\left(\frac{m_{1}a_{1}}{r^{2}+a_{1}^{2}}+\frac{m_{2}a_{2}}{r^{2}+a_{2}^{2}}\right)
\nonumber\\
&&\ \ \ \ \ +\frac{(1+g^{2}r^{2})^{2}(r^{2}+a_{1}^{2})^{2}(r^{2}+a_{2}^{2})^{2}}{\Delta_{r}}
\left(\frac{m_{1}a_{1}}{r^{2}+a_{1}^{2}}+\frac{m_{2}a_{2}}{r^{2}+a_{2}^{2}}\right)^{2}\nonumber\\
&&\ \ \ \ \ -\frac{a_{1}^{2}a_{2}^{2}j(j+D-7)}{r^{2}}-b_{1}r^{2}-b_{2}\Bigg] R_{r}=0,
\end{eqnarray}
and the angular equations, for $i=1,2$,
\begin{eqnarray}
&&\left(\frac{a_{i}}{y_{i}}\right)^{D-6}\frac{d}{dy_{i}}\left[\left(\frac{y_{i}}{a_{i}}\right)^{D-6}(1-g^{2}y_{i}^{2})(a_{1}^{2}-y_{i}^{2})
(a_{2}^{2}-y_{i}^{2})\frac{dR_{y_{i}}}{dy_{i}}\right]\nonumber\\
&&\ \ +\Bigg\{-\frac{(a_{1}^{2}-y_{i}^{2})(a_{2}^{2}-y_{i}^{2})\omega^{2}}{1-g^{2}y_{i}^{2}}
+2\omega[m_{1}a_{1}(a_{2}^{2}-y_{i}^{2})+m_{2}a_{2}(a_{1}^{2}-y_{i}^{2})]\nonumber\\
&&\ \ \ \ \ \ \ -2m_{1}a_{1}m_{2}a_{2}(1-g^{2}y_{i}^{2})-\frac{m_{1}^{2}a_{1}^{2}(1-g^{2}y_{i}^{2})(a_{2}^{2}-y_{i}^{2})}
{a_{1}^{2}-y_{i}^{2}}\nonumber\\
&&\ \ \ \ \ \ \ -\frac{m_{2}^{2}a_{2}^{2}(1-g^{2}y_{i}^{2})(a_{1}^{2}-y_{i}^{2})}
{a_{2}^{2}-y_{i}^{2}}-\frac{a_{1}^{2}a_{2}^{2}j(j+D-7)}{y_{i}^{2}}
-b_{1}y_{i}^{2}+b_{2}\Bigg\}R_{y_{i}}=0.\nonumber\\ \label{angulareqns}
\end{eqnarray}

In the following we shall solve the coupled angular equations above for the case with two rotations to obtain the angular eigenvalues $b_{1}$ and $b_{2}$. Assuming that the rotation parameters $a_{1}$ and $a_{2}$ are small, we shall express these eigenvalues perturbatively as power series of $a_{1}$ and $a_{2}$. However, the case with $D=5$ is exceptional in which there is only one angular equation which has been dealt with recently~\cite{CCDN}. Therefore, we shall consider in the next section the case with $D\geq 6$.

\section{Angular equation and eigenvalue expansion for $D\geq 6$}

With two rotations there are two angular equations involving the latitude coordinates $y_{1}$ and $y_{2}$ as in Eq.~(\ref{angulareqns}) for dimensions $D\geq 6$. In this section we develop expansions in powers of the rotation parameters $a_{1}$ and $a_{2}$ for the angular eigenvalues. Since the latitude coordinates are restricted to $a_{2}\leq y_{1}\leq a_{1}$ and $0\leq y_{2}\leq a_{2}$, it is convenient to change variables to $x_{1}$ and $x_{2}$ with
\begin{eqnarray}
y_{1}^{2}=\frac{1}{2}\left(a_{1}^{2}+a_{2}^{2}\right)-\frac{1}{2}\left(a_{1}^{2}-a_{2}^{2}\right)x_{1}\ \ \ \ \ ;\ \ \ \ \ y_{2}^{2}=\frac{1}{2}a_{2}^{2}\left(1-x_{2}\right),
\end{eqnarray}
where $-1\leq x_{1},x_{2}\leq 1$.

To further simplify the consideration we shall specialize to the case with $D=6$, $j=0$, and $m_{1}=m_{2}=1$. The analysis for other values of these parameters can be carried out in a similar fashion. In this special case the angular equations in Eq.~(\ref{angulareqns}) are ${\cal O}_{i}R_{i}=B_{i}R_{i}/4$ with $i=1,2$. Here, the operators are
\begin{eqnarray}
{\cal O}_{1}
&=&-\frac{(1+x_{1})}{2}\left[(1-x_{1})+\epsilon^{2}(1+x_{1})\right]\left\{2-g^{2}a_{1}^{2}\left[(1-x_{1})+\epsilon^{2}(1+x_{1})\right]
\right\}\frac{d^{2}}{dx_{1}^{2}}\nonumber\\
&&\ \ +\left\{\left[\frac{1}{2}(1+5x_{1})-\frac{1}{4}g^{2}a_{1}^{2}(1-x_{1})(3+7x_{1})\right]\right.\nonumber\\
&&\ \ \ \ \ \ \ \ \ \ \ \left.-\frac{\epsilon^{2}}{2}\left(\frac{1+x_{1}}{1-x_{1}}\right)\left[(1-5x_{1})+7g^{2}a_{1}^{2}x_{1}(1-x_{1})\right]\right.\nonumber\\
&&\ \ \ \ \ \ \ \ \ \ \ \left.+\frac{g^{2}a_{1}^{2}\epsilon^{4}(1+x_{1})^{2}(3-7x_{1})}{4(1-x_{1})}\right\}\frac{d}{dx_{1}}\nonumber\\
&&\ \ +\left\{\frac{a_{1}^{2}\omega^{2}(1+x_{1})(1-\epsilon^{2})^{2}}{4[2-g^{2}a_{1}^{2}((1-x_{1})+\epsilon^{2}(1+x_{1}))]}\right.\nonumber\\
&&\ \ \ \ \ \ \ \ \ \ \left.+\frac{[(1-x_{1})^{2}+\epsilon^{2}(1+x_{1})^{2}][2-g^{2}a_{1}^{2}((1-x_{1})+\epsilon^{2}(1+x_{1}))]}{4(1-x_{1})^{2}(1+x_{1})}\right.\nonumber\\
&&\ \ \ \ \ \ \ \ \ \ \ \ \left.-\frac{\epsilon^{2}(1+x_{1})B_{1}}{4(1-x_{1})}+\frac{B_{2}}{2(1-x_{1})}\right\},\\
{\cal O}_{2}
&=&-\frac{1}{4}(1-x_{2}^{2})\left[2-\epsilon^{2}(1-x_{2})\right]\left[2-g^{2}a_{1}^{2}\epsilon^{2}(1-x_{2})\right]\frac{d^{2}}{dx_{2}^{2}}\nonumber\\
&&\ \ -\left\{\frac{1}{2}(1-3x_{2})+\frac{1}{4}\epsilon^{2}(1+g^{2}a_{1}^{2})(1-x_{2})(1+5x_{2})\right.\nonumber\\
&&\ \ \ \ \ \ \ \ \ \ \left.-\frac{1}{8}g^{2}a_{1}^{2}\epsilon^{4}(1-x_{2})^{2}(3+7x_{2})\right\}\frac{d}{dx_{2}}\nonumber\\
&&\ \ +\left\{\frac{\omega^{2}a_{1}^{2}\epsilon^{2}(1+x_{2})(2-\epsilon^{2}(1-x_{2}))}{8(2-g^{2}a_{1}^{2}\epsilon^{2}(1-x_{2}))}\right.\nonumber\\
&&\ \ \ \ \ \ \ \ \ \ \left.+\frac{[2-g^{2}a_{1}^{2}\epsilon^{2}(1-x_{2})][4-\epsilon^{2}(3-6x_{2}-x_{2}^{2})+\epsilon^{4}(1-x_{2})^{2}]}
{8(1+x_{2})(2-\epsilon^{2}(1-x_{1}))}\right.\nonumber\\
&&\ \ \ \ \ \ \ \ \ \ \left.+\frac{\epsilon^{2}(1-x_{2})B_{1}}{8}\right\},
\end{eqnarray}
where we have defined $\epsilon=a_{2}/a_{1}$. Since we have assumed that $a_{1}\geq a_{2}$, $\epsilon\leq 1$. Moreover, we have defined the constants
\begin{eqnarray}
B_{1}&\equiv&b_{1}+2\omega(a_{1}+a_{2})-2g^{2}a_{1}a_{2},\\
B_{2}&\equiv&\frac{1}{a_{1}^{2}}\left[b_{2}+2\omega(a_{1}a_{2}^{2}+a_{1}^{2}a_{2})-2a_{1}a_{2}\right].
\end{eqnarray}
With these definitions of $B_{1}$ and $B_{2}$, one has only even powers of $\epsilon$ and $a_{1}$.

We shall develop perturbative expansions for $B_{1}$ and $B_{2}$. Since we have assumed that $a_{1}\geq a_{2}$, we first expand the operators with respect to $\epsilon$. Schematically we write
\begin{eqnarray}
&&\left({\cal O}_{i0}+{\cal O}_{i2}+{\cal O}_{i4}+{\cal O}_{i6}+\cdots\right)\left(R_{i0}+R_{i2}+R_{i4}+R_{i6}+\cdots\right)\nonumber\\
&=&\frac{1}{4}\left(B_{i0}+B_{i2}+B_{i4}+B_{i6}+\cdots\right)\left(R_{i0}+R_{i2}+R_{i4}+R_{i6}+\cdots\right).\label{expansion}
\end{eqnarray}
To the zeroth order of $\epsilon$, the eigenvalue equations are ${\cal O}_{i0}R_{i0}=B_{i0}R_{i0}/4$ where
\begin{eqnarray}
{\cal O}_{10}&=&
-\frac{(1-x_{1}^{2})}{2}\left[2-g^{2}a_{1}^{2}(1-x_{1})
\right]\frac{d^{2}}{dx_{1}^{2}}\nonumber\\
&&\ \ +\left[\frac{1}{2}(1+5x_{1})-\frac{1}{4}g^{2}a_{1}^{2}(1-x_{1})(3+7x_{1})\right]
\frac{d}{dx_{1}}\nonumber\\
&&\ \ +\left\{\frac{a_{1}^{2}\omega^{2}(1+x_{1})}{4[2-g^{2}a_{1}^{2}(1-x_{1})]}
+\frac{[2-g^{2}a_{1}^{2}(1-x_{1})]}{4(1+x_{1})}+\frac{B_{20}}{2(1-x_{1})}\right\},\\
{\cal O}_{20}
&=&-(1-x_{2}^{2})\frac{d^{2}}{dx_{2}^{2}}-\frac{1}{2}(1-3x_{2})\frac{d}{dx_{2}}
+\frac{1}
{2(1+x_{2})}.
\end{eqnarray}
Since ${\cal O}_{20}$ does not involve $a_{1}$, the corresponding eigenvalue equation can be solved readily. To fix notation and to anticipate the analysis in the following we define ${\cal O}_{200}\equiv{\cal O}_{20}$. The first zero in the subscript indicates that the operator involves zeroth power of $\epsilon$ while the second zero indicates that it involves zeroth power of $a_{1}$. The eigenfunctions and eigenvalues of the equation ${\cal O}_{200}R_{200}=B_{200}R_{200}/4$ are
\begin{eqnarray}
R_{200}(n_{2})&=&\sqrt{\frac{(4n_{2}+3)(2n_{2}+1)}{2^{3/2}(n_{2}+1)}}(1+x_{2})^{1/2}P^{(-\frac{1}{2},1)}_{n_{2}}(x_{2}),\\
B_{200}(n_{2})&=&2(n_{2}+1)(2n_{2}+1),
\end{eqnarray}
where $n_{2}=0,1,2,\dots$ and $P^{\alpha,\beta}_{n}(x)$ is the Jacobi polynomial. One can check that the orthonormal condition is
\begin{eqnarray}
\frac{1}{4}\int_{-1}^{1}\frac{dx_{2}}{\sqrt{1-x_{2}}}R_{200}(n_{2})R_{200}(n'_{2})=\delta_{n_{2}n'_{2}}.
\end{eqnarray}
To be concrete in the following discussion we take $n_{2}=0$. Hence, we have
\begin{eqnarray}
R_{200}&=&R_{200}(0)=2^{-3/4}3^{1/2}\sqrt{1+x_{2}},\\
B_{200}&=&B_{200}(0)=2.
\end{eqnarray}

Next, we work on the ${\cal O}_{10}$ equation. Following what we have been doing we expand the eigenvalue equation in powers of $a_{1}$ with the operators
\begin{eqnarray}
{\cal O}_{100}&=&-(1-x_{1}^{2})\frac{d^{2}}{dx_{1}^{2}}+\frac{1}{2}(1+5x_{1})\frac{d}{dx_{1}}+\frac{3+x_{1}}{2(1-x_{1}^{2})},\\
{\cal O}_{102}&=&\frac{1}{2}g^{2}a_{1}^{2}(1-x_{1})(1-x_{1}^{2})\frac{d^{2}}{dx_{1}^{2}}-\frac{1}{4}g^{2}a_{1}^{2}(1-x_{1})(3+7x_{1})\frac{d}{dx_{1}}
\nonumber\\
&&\ \ +\frac{[\omega^{2}a_{1}^{2}(1+x_{1})^{2}-2g^{2}a_{1}^{2}(1-x_{1})]}{8(1+x_{1})},\\
{\cal O}_{104}&=&\frac{1}{16}g^{2}\omega^{2}a_{1}^{4}(1-x_{1}^{2}),\\
{\cal O}_{106}&=&\frac{1}{32}g^{4}\omega^{2}a_{1}^{6}(1-x_{1})(1-x_{1}^{2}).
\end{eqnarray}
Similarly we solve the zeroth order eigenvalue equation, ${\cal O}_{100}R_{100}=B_{100}R_{100}/4$, to give
\begin{eqnarray}
R_{100}(n_{1})&=&\sqrt{\frac{(4n_{1}+7)(2n_{1}+5)}{2^{7/2}(n_{1}+1)}}\sqrt{1-x_{1}^{2}}P_{n_{1}}^{(\frac{3}{2},1)}(x_{1}),\\
B_{100}(n_{1})&=&2(n_{1}+1)(2n_{1}+5),
\end{eqnarray}
with $n_{1}=0,1,2,\dots$ and the orthonormal condition
\begin{eqnarray}
\frac{1}{4}\int_{-1}^{1}dx_{1}\sqrt{1-x_{1}}\ R_{100}(n_{1})R_{100}(n'_{1})=\delta_{n_{1}n'_{1}}.
\end{eqnarray}
Again for simplicity we choose $n_{1}=0$ and we have
\begin{eqnarray}
R_{100}&=&R_{100}(0)=2^{-7/4}5^{1/2}7^{1/2}\sqrt{1-x_{1}^{2}},\\
B_{100}&=&B_{100}(0)=10.
\end{eqnarray}
To obtain $B_{102}$ to $B_{106}$, we follow a perturbative procedure~\cite{CCDN}. The result is
\begin{eqnarray}
B_{102}&=&\frac{2}{9}\left(2\omega^{2}a_{1}^{2}-17g^{2}a_{1}^{2}\right),\\
B_{104}&=&-\frac{20}{8019}\left(\omega^{4}a_{1}^{4}-53g^{2}\omega^{2}a_{1}^{4}+196g^{4}a_{1}^{4}\right),\\
B_{106}&=&\frac{20}{8444007}\left(2\omega^{6}a_{1}^{6}-645g^{2}\omega^{4}a_{1}^{6}+28959g^{4}\omega^{2}a_{1}^{6}-105644g^{6}a_{1}^{6}\right).
\end{eqnarray}

If one iterates to higher orders, one can in principle obtain the eigenvalues in powers of $a_{1}$ and $a_{2}=a_{1}\epsilon$. Here we just list the result. For $D=6$, $j=0$, $m_{1}=m_{2}=1$ and $n_{1}=n_{2}=0$,
\begin{eqnarray}
B_{1}&=&\left[10+\frac{2}{9}\left(2\omega^{2}-17g^{2}\right)a_{1}^{2}-\frac{20}{8019}(\omega^{4}-53g^{2}\omega^{2}+196g^{4})a_{1}^{4}\right.\nonumber\\
&&\ \ \ \ \ \ \ \ \ \ \left.+\frac{20}{8444007}(2\omega^{6}-645g^{2}\omega^{4}+28959g^{4}\omega^{2}-105644g^{6})a_{1}^{6}\right]\nonumber\\
&&+\epsilon^{2}\left[\frac{2}{9}(2\omega^{2}-17g^{2})a_{1}^{2}+\frac{32}{8019}(\omega^{4}-53g^{2}\omega^{2}+196g^{4})a_{1}^{4}\right.\nonumber\\
&&\ \ \ \ \ \ \ \ \ \ \ \left.
-\frac{32}{2814669}(\omega^{6}-147g^{2}\omega^{4}+5178g^{4}\omega^{2}-18424g^{6})a_{1}^{6}\right]\nonumber\\
&&+\epsilon^{4}\left[-\frac{20}{8019}(\omega^{4}-53g^{2}\omega^{2}+196g^{4})a_{1}^{4}\right.\nonumber\\
&&\ \ \ \ \ \ \ \ \ \ \ \left.-\frac{32}{2814669}(\omega^{6}-147g^{2}\omega^{4}+5178g^{4}\omega^{2}-18424g^{6})a_{1}^{6}\right]\nonumber\\
&&+\epsilon^{6}\left[\frac{20}{8444007}(2\omega^{6}-645g^{2}\omega^{4}+28959g^{4}\omega^{2}-105644g^{6})a_{1}^{6}\right]+\cdots,
\end{eqnarray}
\begin{eqnarray}
B_{2}&=&2+\epsilon^{2}\left[2+\frac{2}{9}(4\omega^{2}-7g^{2})a_{1}^{2}-\frac{4}{8019}(\omega^{4}-53g^{2}\omega^{2}+196g^{4})a_{1}^{4}\right.\nonumber\\
&&\ \ \ \ \ \ \ \ \ \ \ \ \ \left.+\frac{4}{8444007}(2\omega^{6}-645g^{2}\omega^{4}+28959g^{4}\omega^{2}-105644g^{6})a_{1}^{6}\right]\nonumber\\
&&+\epsilon^{4}\left[-\frac{4}{8019}(\omega^{4}-53g^{2}\omega^{2}+196g^{4})a_{1}^{4}\right.\nonumber\\
&&\ \ \ \ \ \ \ \ \ \ \ \ \left.-\frac{32}{8444007}(4\omega^{2}-25g^{2})(\omega^{4}-53g^{2}\omega^{2}+196g^{4})a_{1}^{6}\right]\nonumber\\
&&+\epsilon^{6}\left[\frac{4}{8444007}(2\omega^{6}-645g^{2}\omega^{4}+28959g^{4}\omega^{2}-105644g^{6})a_{1}^{6}\right]+\cdots.
\end{eqnarray}

\section{Conclusions}\label{sec:conclusion}

In this paper we have started to study higher-dimensional Kerr-(A)dS black holes with two rotations. First, we have presented here the general metric for these doubly rotating black holes. We have also separated the Klein-Gordon equation, writing out the corresponding radial and angular equations explicitly. With these separated equations we could start to ask questions about stability, the various spectra of Hawking radiation and the phenomenon of superradiance.

To get some quantitative understanding of the angular equations, we have developed perturbative expansions in powers of the rotation parameters $a_{1}$ and $a_{2}$ for the angular eigenvalues. These expansions could be used in a variety of considerations. For example, in the evaluation of the frequencies of the low-lying quasinormal modes using a semi-analytic method like the WKB approximation~\cite{iyer}, it is necessary to keep the frequency $\omega$ arbitrary. Perturbative expansions of the angular eigenvalues that we have developed would be useful. In fact, we have already considered the quasinormal modes of these doubly rotating black holes using a numerical approach recently~\cite{CDNC}.

On the other hand, to discuss the stability of the ultra-spinning black holes, we need to consider the angular eigenvalues in the large rotation limit \cite{berti}. For our case there are two rotation parameters. With one small and one large, the situation will be very much like the simply rotating case \cite{morisawa,cardoso} and no instability is expected. Therefore it is more interesting to consider the other case with both rotation parameters large. This work will be pursued subsequently.

\section*{Acknowledgments}

HTC was supported in part by the National Science Council of the Republic of China
under the Grant NSC 99-2112-M-032-003-MY3, and the National Science Centre for
Theoretical Sciences. The work of JD was supported by the
Japan Society for the Promotion of Science (JSPS), under fellowship no. P09749.


\end{document}